# High-Throughput Screening of Transition Metal-Based 2D Multilayer Kagome Materials via the "1 + 3" Design Strategy

Xing-Yu Wang,[1,2] En-Qi Bao,[1] Su-Yang Shen,[1] Jun-Hui Yuan,[1,*] and Jiafu Wang,[1,2]

[1]School of Physics and Mechanics, Wuhan University of Technology, Wuhan 430070, China

[2]School of Materials and Microelectronics, Wuhan University of Technology, Wuhan 430070, China

[*]**Corresponding Author**

E-mail: yuanjh90@163.com (J.-H. Yuan)

## Abstract

Two-dimensional (2D) kagome materials have drawn extensive research interest due to their unique electronic properties, like flat bands, magnetic frustration, and topological quantum states, which enable precise quantum state control and novel device innovation. Yet, simultaneously achieving high stability, tunability, and multifunctionality in 2D kagome systems remains a key material design challenge. In this study, we innovatively propose a new paradigm for constructing two-dimensional multi-kagome-layer materials based on the "1+3" design concept. By seamlessly integrating high-throughput screening techniques, we have successfully identified 6,379 novel 2D multilayer kagome candidates from a vast pool of candidates. These materials exhibit a rich diversity of types, encompassing 173 metals, 27 semimetals, 166 ferromagnetic semiconductors, and as many as 6,013 semiconductors. Furthermore, based on the 2D flat-band scoring criteria, we conducted a detailed analysis of the flat-band characteristics of the energy bands near the Fermi level in the predicted systems. Our findings reveal that approximately two-thirds of the systems meet the 2D flat-band scoring criteria, and notably, several systems exhibit nearly perfect flat-band characteristics. Our work provides an excellent paradigm for the design and research of 2D multilayer kagome materials.

**Keywords:** High-throughput computing, 2D kagome materials, flat bands, semiconductor, first-principles calculations.

## 1. Introduction

In the forefront of condensed matter physics, the kagome lattice has attracted significant attention due to its unique geometric frustration and electronic structural characteristics.[1–5] Its band structure features Dirac cones, van Hove singularities, and localized flat bands, providing a theoretical platform for inducing quantum phenomena such as ferromagnetism and the fractional quantum Hall effect.[6–9] However, intrinsic kagome electronic properties are scarce in real materials, with only about 7% of known kagome network-structured materials exhibiting relevant electronic characteristics.[10] This scarcity stems from the conflict between the two-dimensional (2D) nature of the kagome lattice and the traditional research paradigm of three-dimensional (3D) bulk materials, where key electronic bands in three-dimensional systems are often obscured or shifted away from the Fermi level, making it difficult to observe and utilize novel physical properties.[11,12]

To overcome these bottlenecks, constructing a "pure" 2D kagome material system is an inevitable choice. One of the mainstream strategies involves preparing monolayer counterparts of bulk kagome materials.[13,14] Although these have demonstrated novel physical properties, experimental synthesis poses significant challenges, with few successful cases.[15,16] Moreover, single-layer kagome systems have limitations in the deep modulation of quantum states. In contrast, multilayer kagome systems offer superior modulation potential.[17–19] Theoretical calculations indicate that interlayer coupling in these systems can induce multiple van Hove singularities, giving rise to novel quantum states, while structural flexibility provides abundant degrees of freedom for band modulation.[20,21] However, current research in this field has primarily focused on single-layer kagome systems, with studies on 2D bilayer or multilayer kagome materials still in their infancy. This area faces two core challenges: first, a scarcity of material systems, lacking a systematic database of multi-kagome-layer materials and efficient, universal theoretical design strategies; second, an unclear structure-property relationship, where the quantitative correlation between interlayer coupling strength and electronic state evolution has not been established, hindering the design of quantum materials.

Traditional methods for achieving intrinsic 2D kagome structures suffer from poor scalability, making it difficult to meet the demands of high-throughput material exploration. The innovative "1+3" design paradigm offers a new approach. By utilizing triangular lattice building blocks of 2D materials and introducing symmetry breaking within a 2×2 supercell, this strategy induces spontaneous reconstruction of the system into a kagome lattice. This can be achieved through various local modulation techniques, demonstrating high flexibility and universality. It has been validated in multiple systems and extended to metal oxide monolayers.[22] Based on the "1+3" strategy, the design of single-layer kagome materials has adopted a "skeleton + passivation" model. However, when applied to bilayer kagome systems, complexity increases, requiring consideration of the synergistic effects between interlayer stacking modes and passivation sites. Two-dimensional bilayer kagome structures are primarily classified into Type-I (AA stacking with $C_6$ symmetry)[23] and Type-II (rotated-translated stacking with $C_3$ symmetry),[24] as shown in **Figure S1**. Theoretical studies have predicted some candidates, but overall research remains in its nascent stages, particularly lacking systematic exploration of transition metal-based Type-II systems. Due to their unique combination of symmetries, Type-II bilayer kagome systems exhibit exceptional scientific value, serving as an excellent platform for studying many-body effects and topological transport in kagome physics.

Therefore, developing universal design strategies to construct novel 2D multilayer kagome materials and uncovering the intrinsic relationship between interlayer coupling and quantum physical properties are key directions. Building on the experience of the "1+3" strategy, we extend its application to the theoretical design and property research of Type-II transition metal-based 2D multilayer kagome systems. Utilizing first-principles calculations combined with high-throughput screening, we systematically screened 20,160 potential combinations of 2D transition metal-based multilayer kagome materials. Through stability analysis, we identified 6,379 novel stable 2D multilayer kagome materials, classified their electronic structures, and further screened for 2D kagome materials with flat-band characteristics based on 2D flat-band scoring criteria.

## 2. "1+3" Design Strategy

Taking the widely studied $Bi_2Se_3$ crystal structure (**Figure S2**) as an example, we elucidate the design concept for 2D multilayer kagome materials proposed in this work. First, the 2D $Bi_2Se_3$ monolayer is expanded into a 2×2×1 supercell, forming the $Bi_8Se_{12}$ structure. For simplicity of description, this structure is generalized to the $M_8X_{12}$ configuration, as shown in **Figure 1a**. $M_8X_{12}$ consists of two M atomic layers, with M atoms in each layer arranged in a closely packed hexagonal grid. Building on this, following the "1+3" strategy—which involves simultaneously removing M atoms located at the coordinates (2/3, 1/3, up) and (1/3, 2/3, down) in both the upper and lower M atomic layers while retaining three-quarters of the M atoms—the $M_6X_{12}$ crystal configuration (referred to as the "6+12" configuration) can be obtained. This configuration contains two kagome lattices formed by M atoms, resulting in a 2D bilayer kagome configuration. Notably, in the $M_6X_{12}$ structure, if the same "1+3" strategy is applied to the four X sites in the central layer by removing the edge X atoms, the remaining three X atoms precisely form a kagome lattice. In this case, $M_6X_{12}$ can evolve into $M_6X_{11}$ (referred to as the "6+11" configuration). $M_6X_{11}$ contains three kagome layers, including two kagome layers formed by M metal atoms and one kagome layer formed by nonmetal X atoms.

Considering the symmetry of the X sites in $M_6X_{12}$ and $M_6X_{11}$, further potential configurations can be derived. For example, in $M_6X_{12}$, there are four non-equivalent X sites, labeled as X1, X2, X3, and X4 (as shown in **Figure 1**). Introducing different elements to occupy these four non-equivalent sites yields various configurations: when all four X sites are occupied by different elements, the $M_6(X1)_1(X2)_2(X3)_3(X4)_6$ configuration can be formed, referred to as the "6+1+2+3+6" system. Based on the elemental differences occupying the four X sites, the "6+1+2+3+6" system can be further subdivided into eight potential configurations: "6+1+5+6", "6+2+4+6", "6+3+3+6", "6+6+6", "6+1+11", "6+2+10", "6+3+9", and "6+4+8". For $M_6X_{11}$, there are three non-equivalent X sites (X2, X3, X4), with the general configuration being $M_6(X2)_2(X3)_3(X4)_6$ (referred to as "6+2+3+6"). Depending on the similarities or differences among the X2, X3, and X4 elements, this can be further subdivided into three potential configurations: "6+2+9", "6+3+8", and "6+5+6". The above results demonstrate that by adjusting the elemental types at non-equivalent X sites, various material configurations can be obtained, significantly increasing the system's degrees of freedom and providing more opportunities for subsequent research on kagome properties.

After identifying the two basic crystal configurations, $M_6X_{12}$ and $M_6X_{11}$, and their derivative crystal configurations, the next step is to consider the screening range for M and X atoms. In this work, as shown in **Figure 1b**, for the transition metal M, Group IIIB elements Sc and Y, Group IVB elements Ti, Zr, and Hf, and Group VB elements V, Nb, and Ta—a total of eight metals—were examined. For nonmetal coordination at the X sites, Group IVA elements C and Si, Group VA elements N and P, Group VIA elements S, Se, and Te, and Group VIIA elements Cl, Br, and I—a total of ten non-metal elements—were considered. Additionally, during the screening process, there were differences in element selection for the four X sites: the X1, X3, and X4 sites were filled with six elements—S, Se, Te, Cl, Br, and I; the X2 site allowed for filling with ten elements from Group IVA (C, Si), Group VA (N, P), Group VIA (S, Se, Te), and Group VIIA (Cl, Br, I). This established the entire material screening space, totaling 20,160 candidate materials to be screened.

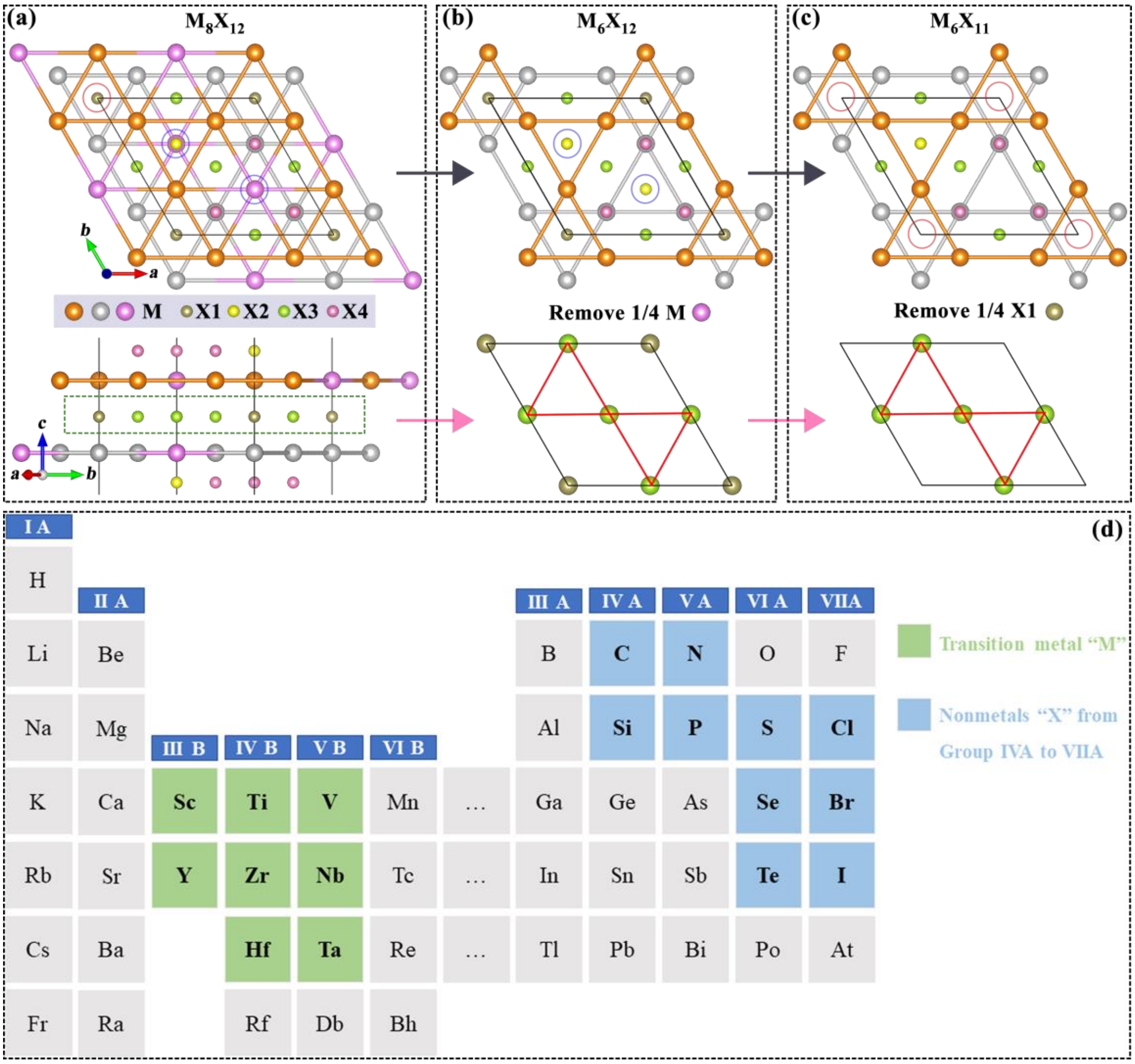

**Figure 1**. Design concept and screening scope for 2D multilayer kagome lattices: (a) Crystal structure of $M_8X_{12}$. (b) Crystal structure of $M_6X_{12}$ after removing 1/4 of the M atoms. (c) Crystal structure of $M_6X_{11}$ after removing 1/4 of the X atoms located at the vertices of the central layer. (d) Element screening scope, where the green section indicates the element screening range for M atoms, while the blue areas represent the element screening ranges for the four nonmetal sites, X1, X2, X3, and X4.

## 3. Result and Discussion

### 3.1 High-Throughput Screening Process

Based on the aforementioned design strategy, we combined first-principles calculations with high-throughput screening to evaluate materials from the $M_6X_{12}$ and $M_6X_{11}$ systems, along with their derivative systems. The screening process is illustrated in **Figure 2**, with the specific steps outlined as follows:

(i). For multiple systems, including "6+12", "6+11", and their derivatives "6+1+2+3+6" and "6+2+3+6", an exhaustive permutation and combination approach was employed using the element sources depicted in **Figure 1b** to generate the database for high-throughput screening.

(ii). Based on density functional theory, structural optimizations and formation energy calculations were performed for all candidate materials using the VASP software. Structures with formation energies less than zero were excluded.

(iii). Materials meeting the formation energy requirements underwent further calculations of phonon spectra, AIMD (at 300 K), and elastic constants to screen for candidates that satisfied kinetic, thermodynamic, and mechanical stability criteria.

(iv). High-throughput band structure calculations were conducted for the finally screened candidates using the hybrid functional HSE06, and the materials were classified based on their band structures (e.g., metals, semimetals, ferromagnetic semiconductors, semiconductors).

(v). Flat-band characteristics were extracted from the band structure calculation results based on 2D flat-band scoring criteria.

For specific parameter settings in the first-principles calculations, please refer to **Note 1** in the **Supplementary Material**. According to the established design strategy and the selected range of screening elements, a total of 20,160 candidate materials were evaluated. As shown in **Figure 2**, after assessing stability criteria, including formation energy calculations, phonon spectra, AIMD, and elastic constants, a total of 6,379 stable systems were identified. Detailed information can be found in **Note 2** of the **Supplementary Material**. Here, "stable" refers to 2D systems that can theoretically exist freely without relying on external conditions such as substrates. Subsequently, the electronic structures of these 6,379 stable candidate materials were calculated using the hybrid functional HSE06. The results indicated that the 6,379 materials could be classified into four categories based on their electronic structures: metals, semimetals, ferromagnetic semiconductors, and semiconductors. Specifically, there were 173 metals, 72 semimetals, 166 ferromagnetic semiconductors, and 6,013 semiconductors. For detailed data, please refer to **Note 3** in the **Supplementary Material**. Next, we present the screening results in detail based on the two material classifications, "6+12" and "6+11".

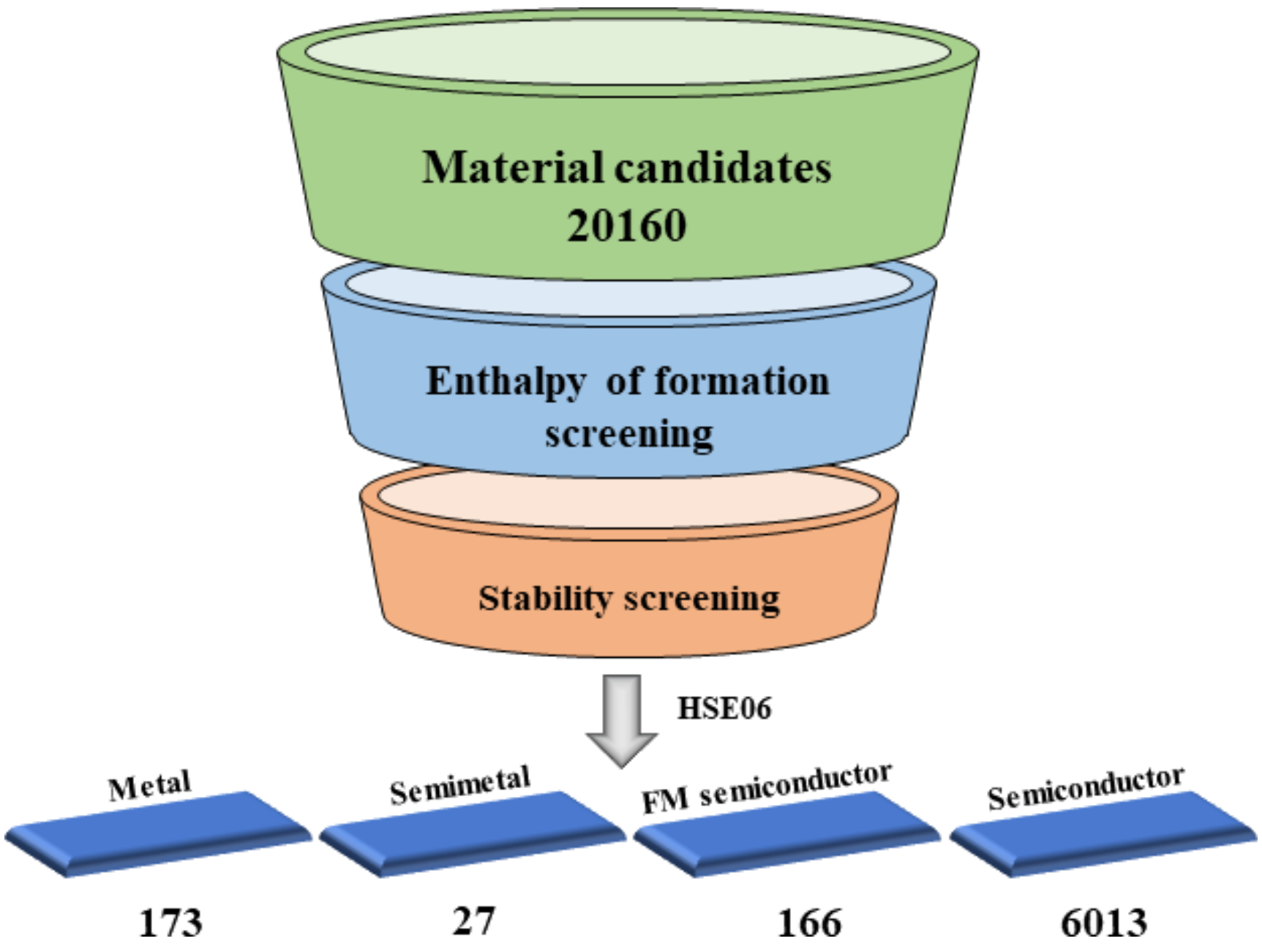

**Figure 2.** Flowchart of the high-throughput screening process. The results currently displayed, especially those concerning the quantities of different types of materials screened, represent preliminary screening outcomes. The final results still require further validation, and we will continue to refine them in subsequent work.

### 3.2 "6+12" System

For the "6+12" system and its derivative systems, their simplest configuration is that of binary compounds. As shown in **Figure 3a**, a total of 15 stable 2D $M_6X_{12}$ compounds have been identified (refer to the **Supplementary Material** for details). Among these, except for $Ti_6Te_{12}$, which exhibits metallic properties, the remaining 14 2D compounds all demonstrate semiconductor characteristics with bandgaps ranging from 0.03 to 1.54 eV. After high-throughput screening, one of the derivative systems of "6+12", namely the "6+6+6" system (as depicted in **Figure 3b**), yielded a total of 76 stable 2D compounds (refer to the **Supplementary Material** for details). Among these, except for $V_6P_6I_6$, which exhibits metallic properties, and $Ta_6S_6Cl_6$ and $Ta_6Te_6I_6$, which exhibit semimetallic properties, the remaining 73 2D compounds all demonstrate semiconductor characteristics. The screening results for other configurations within the "6+12" system are still being processed due to the enormous amount of data involved, and updates will be provided subsequently.

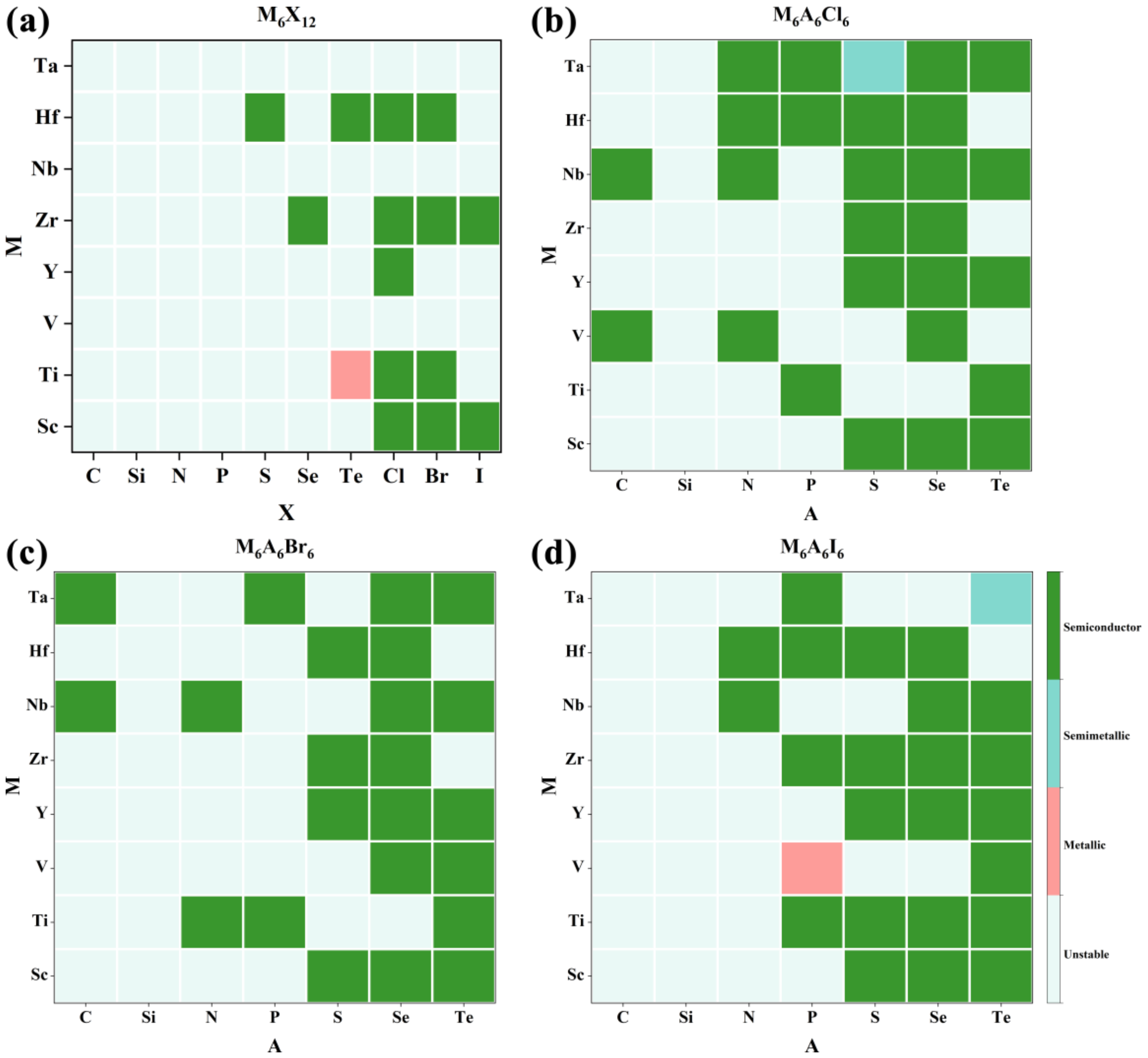

**Figure 3.** Stability screening results for the "6+12" and "6+6+6" systems, where green indicates semiconductor properties, pink indicates metallic properties, light blue indicates semimetallic properties, and light cyan indicates instability.

### 3.3 "6+11" System

Classified by chemical composition, the simplest configuration of the "6+11" system is also a binary compound. After high-throughput stability screening, only one stable 2D structure, $Sc_6Cl_{11}$, was identified in the binary "6+11" system, and it exhibits metallic properties, as shown in **Figure 4a**. Additionally, the "6+11" system can be expanded into the "6+2+3+6" system, which further includes three subclasses: "6+5+6", "6+2+9", and “6+3+8”. For the “6+2+3+6” system, high-throughput screening (as shown in **Figure 4b**) revealed a total of 218 stable compounds (refer to the **Supplementary Material** for details). Among these, 44 2D compounds exhibit metallic properties, while five compounds—$Nb_6Se_2S_3I_6$, $Ti_6S_2Se_3I_6$, $Ti_6Se_2Te_3Br_6$, $Ti_6Se_2Te_3I_6$, and $Y_6Te_2C_3Br_6$—exhibit semimetallic properties. The remaining 169 2D compounds exhibit semiconductor properties. For the “6+5+6” subclass, high-throughput screening, as depicted in **Figure 4c**, identified a total of 69 stable compounds (refer to the **Supplementary Material** for details), with 25 exhibiting metallic properties and 44 exhibiting semiconductor properties.

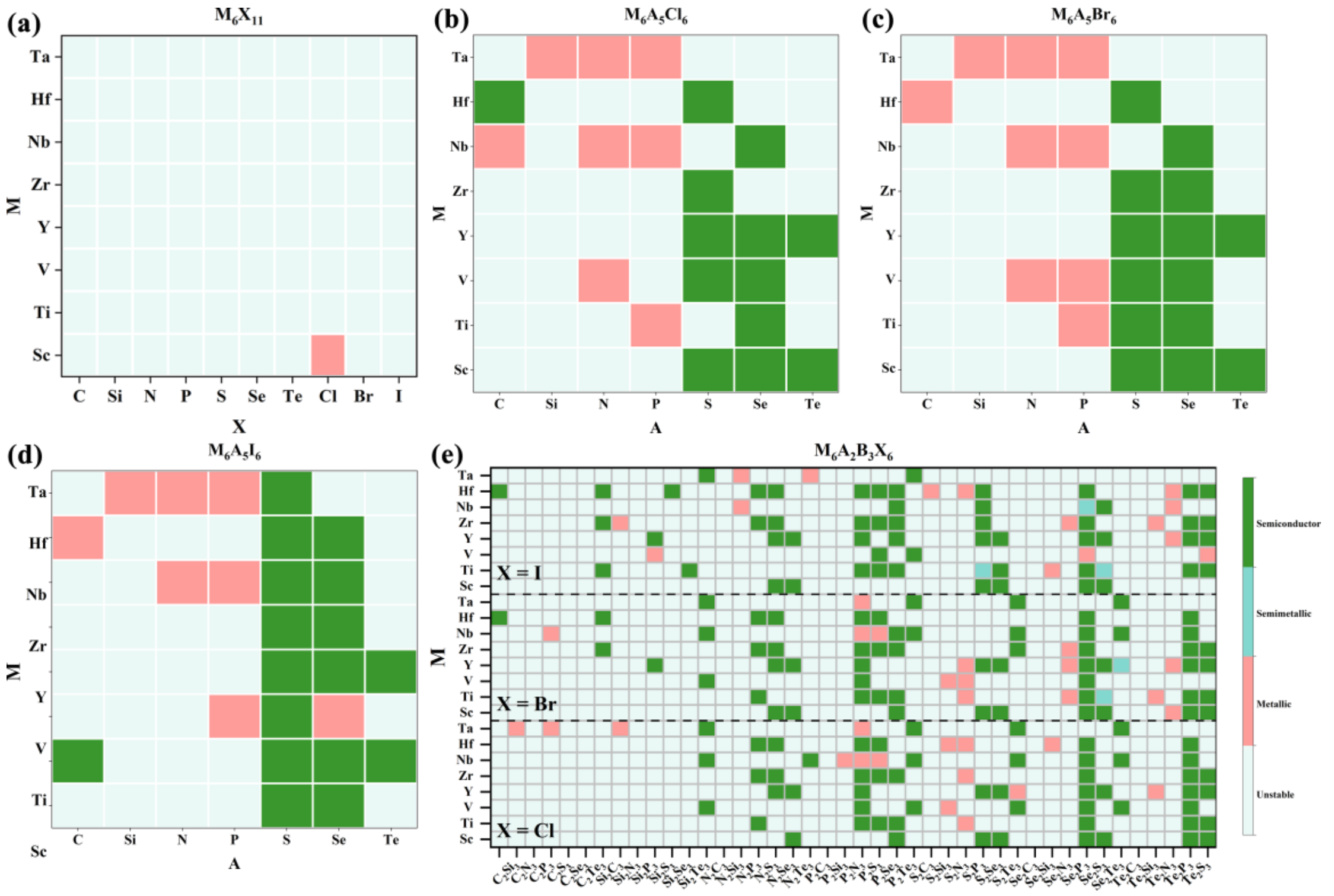

**Figure 4.** (a) Stability screening results for the binary compounds in the “6+11” system; (b-d) Stability screening results for the “6+5+6” system. (b) Stability screening results for the “6+2+3+6” system. In these results, green indicates semiconductor properties, pink indicates metallic properties, light blue indicates semimetallic properties, and light

cyan indicates instability.

### 3.3 Electronic Properties

Based on the aforementioned screening results, we calculated the electronic structures of the stable systems using the hybrid functional HSE06. **Figure 5** displays some representative band structures. The 2D multilayer kagome materials, meticulously designed based on the "1+3" strategy, encompass a rich variety of types from the perspective of electronic structure, including metals, semimetals, semiconductors, and ferromagnetic semiconductors. Taking the "6+12" system as an example, $Ti_6Te_{12}$ within this system exhibits metallic properties (as shown in **Figure 5a**); while the other $M_6X_{12}$ compounds in the system are all semiconductors. For instance, $Hf_6X_{12}$ (X = Cl, Br, I), which possesses a Γ-Γ direct bandgap (as depicted in **Figure 5b**), is particularly notable for having multiple Dirac points at the K point (as clearly shown in **Figure 5b**). As the M element varies, the band structure of $M_6X_{12}$ undergoes significant changes, which are visually demonstrated in **Figure 5c**. Specifically, $Sc_6Cl_{12}$ is an M-M direct bandgap semiconductor; however, when Cl vacancies are introduced into the central Cl atomic layer of $Sc_6Cl_{12}$, forming the $Sc_6Cl_{11}$ structure, its nature undergoes a fundamental transformation, directly changing from a direct bandgap semiconductor to a metal (as shown in **Figure 5d**), and a nearly perfect flat band emerges near approximately 0.6 eV.

In the subsequent "6+6+6" family, most systems within the Sc and Y families well preserve the kagome band characteristics.[25,26] Taking $Y_6S_6Cl_6$ as an example, the energy band at the top of the valence band almost possesses the characteristics of a perfect flat band, while the conduction band also retains relatively complete kagome band features, including flat bands and Dirac cones. After introducing S vacancies, while maintaining the original flat-band characteristics of the valence band unchanged, the kagome bands in the original conduction band undergo decoupling and evolve into isolated bands near the Fermi level. Meanwhile, a large bandgap (> 1 eV) opens up at the original Dirac cone at the K point. $Ta_6S_6Cl_6$, which has the same structure, is a Γ-Γ direct narrow-bandgap semiconductor (with a bandgap of 0.16 eV), similar to $Hf_6Cl_{12}$ (as shown in **Figure 5g**). When Se vacancies are introduced, the bandgap of $Ta_6Se_5Br_6$ increases to 0.31 eV, but the bandgap type changes from a direct bandgap to an indirect bandgap (as seen in **Figure 5h**). The main difference in the electronic structures of the two lies in the valence band, which is mainly contributed by the Se atoms in the central layer.

For the "6+2+3+6" system, it features a rich variety of materials and even more complex and variable band structures, as shown in **Figures 5i-5l**. $Ti_6Se_2Te_3I_6$ is a semimetal with Dirac cones at the K point (as shown in **Figure 5i**). In the band structure of $Y_6N_2Se_3Cl_6$, there are almost perfect flat-band characteristics in the valence band and isolated bands in the conduction band. Its band structure is similar to that of $Y_6S_5Cl_6$, except that $Y_6N_2Se_3Cl_6$ has a lower Fermi level (as shown in **Figure 5j**). All Nb-based compounds in the "6+2+3+6" system are metals. For instance, $Nb_6P_2Se_3Br_6$ is a metal with multiple Dirac points near the Fermi level (as shown in **Figure 5k**). When the elements of Group VA, VIA, and VIIA at the X1 site are replaced with Group IVA elements, magnetism is mostly introduced into the system. Taking $Hf_6C_2S_3Cl_6$ as an example (as shown in **Figure 5l**), it is a ferromagnetic semiconductor, with both spin-up and spin-down states exhibiting semiconductor characteristics, providing a highly promising candidate material for applications in the field of magnetism.

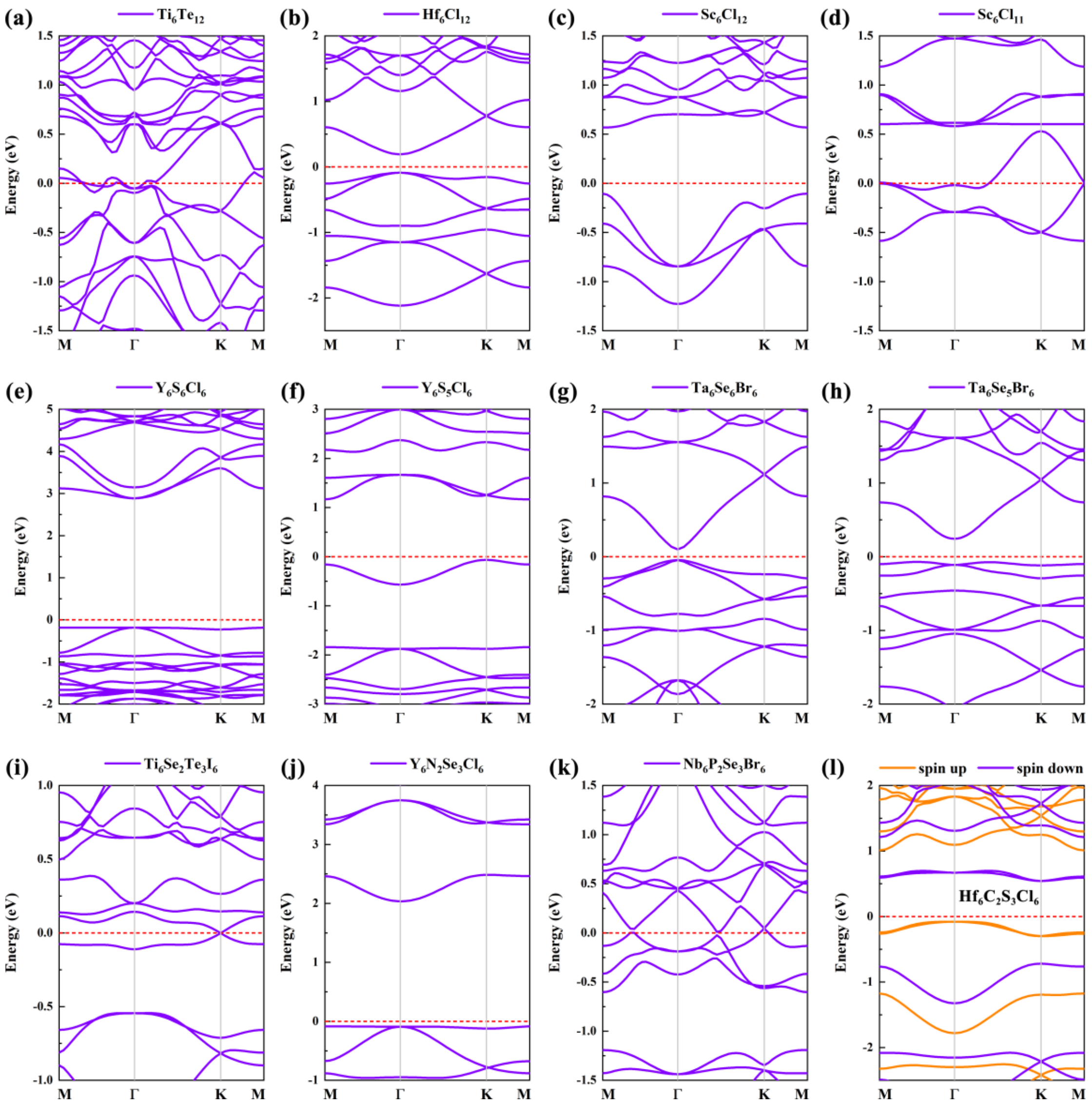

**Figure 5.** The calculated band structure of (a) $Ti_6Te_{12}$, (b) $Hf_6Cl_{12}$, (c) $Sc_6Cl_{12}$, (d) $Sc_6Cl_{11}$, (e) $Y_6S_6Cl_6$, (f) $Y_6S_5Cl_6$, (g) $Ta_6Se_6Br_6$, (h) $Ta_6Se_5Br_6$, (i) $Ti_6Se_2Te_3I_6$, (j) $Y_6N_2Se_3Cl_6$, (k) $Nb_6P_2Se_3Br_6$ and (l)$Hf_6C_2S_3Cl_6$ based on HSE06 calculation.

Furthermore, based on the 2D flat-band scoring criteria, flat-band characteristics are extracted from the band calculation results of 2D compounds exhibiting semiconductor properties.[27] With $\Delta E$ = 0.5 eV and $\omega$ = 25 meV, out of a total of 6,013 semiconductor 2D compounds, 4,189 meet the 2D flat-band scoring criteria. Some of these systems even score as high as 1. For specific scoring criteria of other systems, please refer to the **Supplementary Material**.

## 4. Conclusion

In summary, based on the “1+3” design strategy, we employed atomic ratio design and high-throughput screening to identify a total of 6,379 stable 2D multilayer kagome materials. To further investigate the electronic structures of these stable systems, HSE06 calculations were performed, revealing that the final stable systems comprised 173 metals, 27 semimetals, 166 ferromagnetic semiconductors, and 6,013 semiconductors. Notably, the majority of the 4,189 semiconductors exhibited flat bands. To evaluate their flat-band characteristics, we utilized a 2D flat-band scoring criterion to extract flat-band features from the energy bands near the Fermi level, identifying many systems with perfect flat bands. Additionally, intriguing findings emerged from the electronic structure calculations, such as isolated energy bands. By deeply integrating novel material design strategies with high-throughput computational techniques, this work not only pioneers a new pathway for the exploration and development of 2D kagome materials but also provides robust material support and theoretical foundations for the research and development of next-generation quantum devices.

## Author Contributions

X.-Y. Wang and E.-Q. Bao performed the calculation and data analysis, and wrote the original manuscript. S.-Y. Shen performed the data analysis. J.-H. Yuan conceived the idea, performed the calculation and data analysis, and wrote and revised the manuscript. J. Wang conducted guidance and data analysis. All authors reviewed this manuscript.

## Notes

The authors declare no competing financial interest.

## Statement

Given the extensive scope of material screening in this work and the potential for data deviations during the high-throughput screening process, we are conducting a meticulous, item-by-item verification of all screening results to ensure the accuracy and reliability of the paper's conclusions. Although only a portion of the screening results is currently presented in the manuscript, this will in no way compromise the validity of the design concepts and strategies proposed therein.

## Acknowledgments

This work was supported the Fundamental Research Funds for the Central Universities (WUT: 2024IVA052). Meanwhile, we would like to express our gratitude to the Feima High-Performance Computing Platform at Wuhan University of Technology for providing computing resource support, and to Hongzhiwei (Shanghai) Technology Co., Ltd. for their software and computing resource support.